\documentclass[prl,aps,twocolumn,superscriptaddress,showpacs]{revtex4}
\usepackage{amsmath,amssymb,graphicx}
\usepackage{pxfonts}
\usepackage{graphicx}
\usepackage{color}
\usepackage{float}
\begin{document}

\title{Long-distance decay-less spin transport in indirect excitons in a van der Waals heterostructure}

\author{Zhiwen~Zhou}
\affiliation{Department of Physics, University of California at San Diego, La Jolla, CA 92093, USA}
\author{E.~A.~Szwed}
\affiliation{Department of Physics, University of California at San Diego, La Jolla, CA 92093, USA}
\author{D.~J. Choksy}
\affiliation{Department of Physics, University of California at San Diego, La Jolla, CA 92093, USA}
\author{L.~H.~Fowler-Gerace}
\affiliation{Department of Physics, University of California at San Diego, La Jolla, CA 92093, USA}
\author{L.~V.~Butov} 
\affiliation{Department of Physics, University of California at San Diego, La Jolla, CA 92093, USA}

\begin{abstract}
\noindent
In addition to its fundamental interest, the long-distance spin transport with suppressed spin losses is essential for spintronic devices. However, the spin relaxation caused by scattering of the particles carrying the spin, limits the spin transport. We explored spatially indirect excitons (IXs), also known as interlayer excitons, in van der Waals heterostructures (HS) composed of atomically thin layers of transition-metal dichalcogenides (TMD) as spin carries. TMD HS also offer coupling of spin and valley transport. We observed the long-distance spin transport with the decay distances exceeding 100~$\mu$m and diverging so spin currents show no decay in the HS. With increasing IX density, we observed spin localization, then long-distance spin transport, and then reentrant spin localization, in agreement with the Bose-Hubbard theory prediction for superfluid  and insulating phases in periodic potentials due to moir{\'e} superlattices. The suppression of scattering in exciton superfluid suppresses the spin relaxation and enables the long-distance spin transport. This mechanism of protection against the spin relaxation makes IXs a platform for the realization of long-distance decay-less spin transport.
\end{abstract}

\maketitle

The physics of spin transport include a number of fundamental phenomena, including, in particular, 
the current-induced spin orientation (the spin Hall effect)~\cite{Dyakonov1971, Hirsch1999, Sinova2004, Kato2004}, 
the spin drift, diffusion and drag~\cite{Kikkawa1999, Weber2005, Crooker2005}, 
the quantum spin Hall effect~\cite{Kane2005, Bernevig2006, Konig2007}, and 
the persistent spin helix~\cite{Koralek2009}.
In addition to its fundamental interest, the long-distance spin transport with suppressed spin losses is essential for developing spintronic devices, which may offer advantages in dissipation, size, and speed over charge-based devices~\cite{Awschalom2007}.

IXs in HS can enable the realization of the long-distance spin transfer. IXs are composed of electrons and holes confined in separated layers~\cite{Lozovik1976}. Due to the separation of electron and hole layers, the IX lifetimes can exceed the lifetimes of spatially direct excitons (DXs) by orders of magnitude. Due to their long lifetimes, IXs can cool down below the temperature of quantum degeneracy and form a condensate~\cite{High2012} and can travel over long distances~\cite{Hagn1995}. Travelling particles can transfer the spin state. However, the particle scattering causes fluctuating effective magnetic fields originating from the spin-orbit interaction in noncentrosymmetric materials and, as a result, causes the spin relaxation that limits the spin transfer~\cite{Dyakonov2008}. Therefore, the suppression of scattering in IX condensate can suppress the spin relaxation and allow the long-distance spin transport. In addition, the electron-hole separation in IXs reduces the overlap of the electron and hole wave functions suppressing the spin relaxation due to electron-hole exchange~\cite{Maialle1993}.

IXs can be created in various HS, in particular, in GaAs HS~\cite{Leonard2009, High2013, Violante2015, Finkelstein2017, Leonard2018}, in GaN HS~\cite{Chiaruttini2019}, and in ZnO HS~\cite{Morhain2005}. Since the temperature of quantum degeneracy, which can be achieved for excitons, scales proportionally to the exciton binding energy $E_{\rm X}$~\cite{Fogler2014}, IXs with high $E_{\rm X}$ can form a platform for the realization of high-temperature long-distance spin transport. 

IXs in GaAs HS have low $E_{\rm X} \lesssim 10$~meV~\cite{Zrenner1992, Sivalertporn2012} and the highest $E_{\rm X} \sim 30$~meV for IXs in III-V and II-VI semiconductor HS is achieved 
in ZnO HS~\cite{Morhain2005}. TMD HS enable the realization of excitons with remarkably high 
binding energies~\cite{Geim2013, Ye2014, Chernikov2014, Goryca2019} and $E_{\rm X}$ for IXs
in TMD HS reach hundreds of meV~\cite{Fogler2014, Berman2017, Deilmann2018}. 

TMD HS also give an opportunity to explore spin transport in periodic potentials due to moir{\'e} superlattices. The period of the latter $b \approx a/{\sqrt{\delta \theta^2 + \delta^2}}$ is typically in the 10~nm range ($a$ is the lattice constant, $\delta$ the lattice mismatch, $\delta \theta$ the deviation of the twist angle between the layers from $i\pi/3$, $i$ is an integer)~\cite{Wu2018, Yu2018, Wu2017, Yu2017, Zhang2017a, Rivera2018, Zhang2018, Ciarrocchi2019, Seyler2019, Tran2019, Jin2019, Alexeev2019, Jin2019a, Shimazaki2020, Wilson2021, Gu2022}. 
The moir{\'e} potentials can be affected by atomic reconstruction~\cite{Weston2020, Rosenberger2020, Zhao2023} and by disorder. In addition, due to the coupling of the spin and valley indices in TMD HS~\cite{Xiao2012, Cao2012, Zeng2012, Mak2012}, the spin transport is coupled to the valley transport (therefore, for simplicity, we will use the term 'spin' also for 'spin-valley').

\begin{figure*}
\begin{center}
\includegraphics[width=15cm]{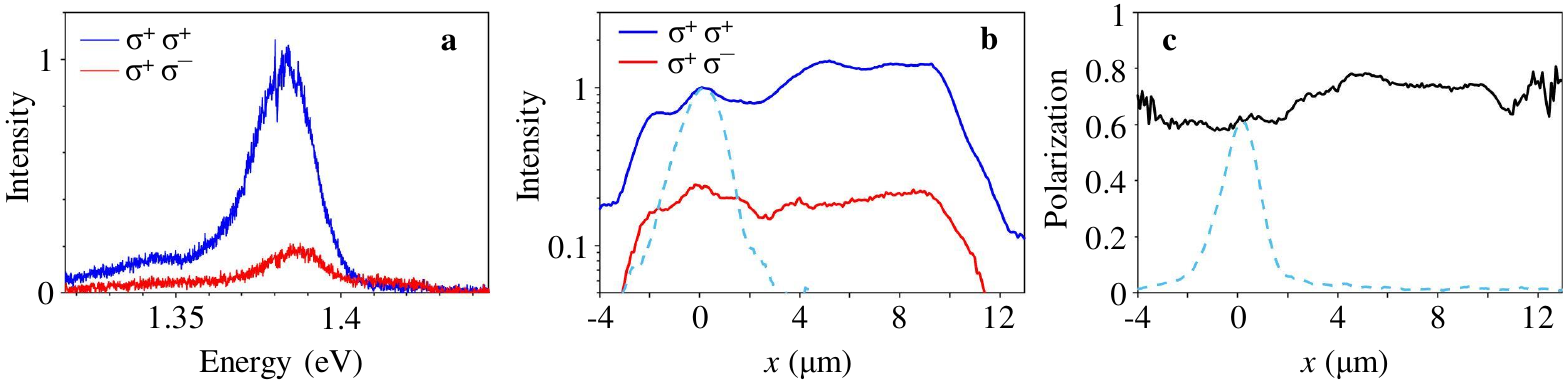}
\caption{The long-distance spin-valley polarization transport in IXs in MoSe$_2$/WSe$_2$ HS. 
(a) The circular polarization of IX PL. The blue (red) spectrum is co-polarized (cross-polarized) with the circularly polarized laser excitation. 
(b) Co-polarized $I_{\sigma^+}$ (blue) and cross-polarized $I_{\sigma^-}$ (red) IX PL intensity vs. the distance from the laser excitation spot centered at $x = 0$. The IX PL is spectrally integrated in the range $E < 1.4$~eV. The dashed line shows the DX PL profile in the MoSe$_2$ ML, this profile is close to the laser excitation profile for short-range DX transport. The HS active area extends from $x \sim -3$ to 10~$\mu$m. The polarized IX PL propagates through the entire HS. 
(c) The degree of circular polarization of IX PL $P = (I_{\sigma^+} - I_{\sigma^-})/(I_{\sigma^+} + I_{\sigma^-})$ vs. the transport distance. No decay is observed for the polarization transport for IXs over the entire HS.
The laser excitation power $P_{\rm ex} = 0.2$~mW, $T = 1.7$~K.
}
\end{center}
\label{fig:spectra}
\end{figure*}

Earlier studies led to the observation of spin transport with $1/e$ decay distances $d^{\rm s}_{1/e}$ up to a few~$\mu$m in IXs in TMD HS~\cite{Rivera2016, Unuchek2019, Huang2020, Shanks2022}. Spin transport on a $\mu$m scale was also observed in DXs~\cite{Onga2017} and the excitation-induced polarization was found to lead to the emergence of ferromagnetic order~\cite{Hao2022} and to electron or hole spin transport with the spin diffusion length up to ca. 20~$\mu$m~\cite{Jin2018, Ren2022} in TMD. Spin relaxation due to scattering of the particles carrying the spin limited spin transport distances~\cite{Dyakonov2008}.

In this work, we observed in IXs in a MoSe$_2$/WSe$_2$ HS the long-distance spin transport with $d^{\rm s}_{1/e}$ exceeding 100~$\mu$m and diverging so spin currents show no decay in the HS. With increasing IX density, we observed spin localization, then long-distance spin transport, and then reentrant spin localization, in agreement with the Bose-Hubbard theory prediction for superfluid and insulating phases in periodic potentials due to moir{\'e} superlattices~\cite{Fisher1989}. The suppression of scattering in exciton superfluid suppresses the spin relaxation and enables the long-distance spin transport.

We study MoSe$_2$/WSe$_2$ HS assembled by stacking mechanically exfoliated 2D crystals [Fig.~S1 in Supplementary Information (SI)]. IXs are formed from electrons and holes confined in adjacent MoSe$_2$ and WSe$_2$ monolayers (ML), respectively, encapsulated by hBN layers. No voltage is applied in the HS. IXs form the lowest-energy exciton state in the MoSe$_2$/WSe$_2$ HS (Fig.~S1 in SI). The HS details are presented in SI.

Both the long-distance IX transport~\cite{Fowler-Gerace2023} and the long-distance spin transport, which is described below, are realized when the optical excitation has the energy $E_{\rm ex}$ close to the energy of DXs in the HS. The resonant excitation allows lowering the excitation-induced heating of the IX system. In particular, the colder IXs created by the resonant excitation screen the HS disorder more effectively that facilitates the emergence of IX superfluidity~\cite{Nikonov1998, Ivanov2006, Remeika2009}. In this work, the laser excitation with $E_{\rm ex} = 1.689$~eV is resonant to WSe$_2$ DX.

\begin{figure*}
\begin{center}
\includegraphics[width=15cm]{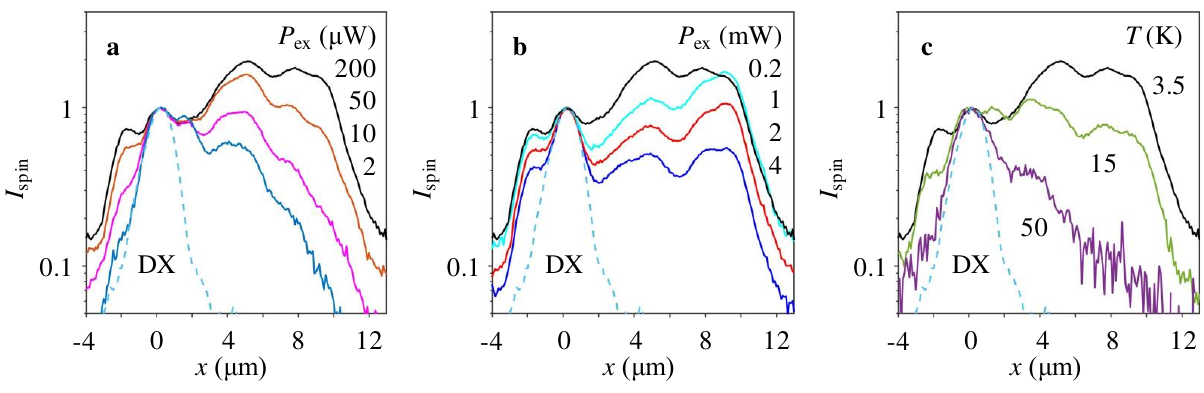}
\caption{Excitation power and temperature dependence of spin transport in LE-IXs. 
(a-c) Normalized spin density profiles $I_{\rm spin} = P n = I_{\sigma^+} - I_{\sigma^-}$ for the LE-IXs for different $P_{\rm ex}$ (a,b) and temperatures (c).
The dashed line shows the DX luminescence profile in the MoSe$_2$ ML, this profile is close to the laser excitation profile for short-range DX transport. 
The LE-IX spectra are separated from the HE-IX spectra by the spectral integration in the range $E < 1.4$~eV. The HE-IXs appear in the spectra at high $P_{\rm ex} \gtrsim 0.2$~mW (Fig.~4). 
The $\sim 2$~$\mu$m laser spot is centered at $x = 0$ (a-c), $T = 3.5$~K (a,b), $P_{\rm ex} = 0.2$~mW (c). 
}
\end{center}
\label{fig:spectra}
\end{figure*}

Both spin generation and detection in IXs is achieved by optical means via photon polarization. The circularly polarized laser excitation is focused to a $\sim 2$~$\mu$m spot and the spin propagation is detected by the polarization-resolved PL imaging. Figure~1a shows a high degree of circular polarization in the excitation spot, indicating an effective relaxation of the optically generated spin-polarized DXs to the spin-polarized IXs with the spin relaxation time long compared to the exciton recombination and energy relaxation times. 

The propagation of spin-polarized IXs from the excitation spot transfer the spin polarization. Remarkably, both the intensities of co-polarized and cross-polarized IX PL $I_{\sigma^+}$ and $I_{\sigma^-}$ (Fig.~1b) and the degree of circular polarization of IX PL $P = (I_{\sigma^+} - I_{\sigma^-})/(I_{\sigma^+} + I_{\sigma^-})$ (Fig.~1c) propagate over the entire HS with no losses. 

IX transport is characterized by the propagation of total IX intensity in both circular polarizations $n \sim I_{\sigma^+} + I_{\sigma^-}$. In turn, transport of spin polarization density carried by IXs is characterized by the propagation of $I_{\rm spin} = P n = I_{\sigma^+} - I_{\sigma^-}$. The dependence of spin transport on excitation power $P_{\rm ex}$ and temperature is described below. 

The spin transport nonmonotonically varies with increasing $P_{\rm ex}$, increases at $P_{\rm ex} \lesssim 0.2$~mW (Fig.~2a) and reduces at $P_{\rm ex} \gtrsim 0.2$~mW  (Fig.~2b), and vanishes at high temperatures (Fig.~2c). The spin transport is characterized by the $1/e$ decay distance of the spin polarization density $d^{\rm s}_{1/e}$. The variation of spin transport with excitation power and temperature is presented by the variation of $d^{\rm s}_{1/e}$ in Fig.~3. 

For low $P_{\rm ex}$, a single IX PL line is observed in the spectra. However, a higher-energy IX PL line appears in the spectrum at high $P_{\rm ex}$ (Fig.~4). We will refer to the IXs corresponding to these PL lines as the lower-energy IXs (LE-IXs) and higher-energy IXs (HE-IXs). Figures 2 and 3 present the spin transport carried by LE-IXs. $d^{\rm s}_{1/e}$ (Fig.~3) are obtained from least-squares fitting the LE-IX spin density transport profiles $I_{\rm spin}(x)$ (Fig.~2) to exponential decays in the HS. The HS dimensions allow establishing that the longest $d^{\rm s}_{1/e}$ exceed 100~$\mu$m. The data with the fit indicating diverging $d^{\rm s}_{1/e}$, that is with no spin density decay within the entire HS, are presented by circles on the edge in Fig.~3a,b and by cyan color in Fig.~3c.

\begin{figure}
\begin{center}
\includegraphics[width=8.5cm]{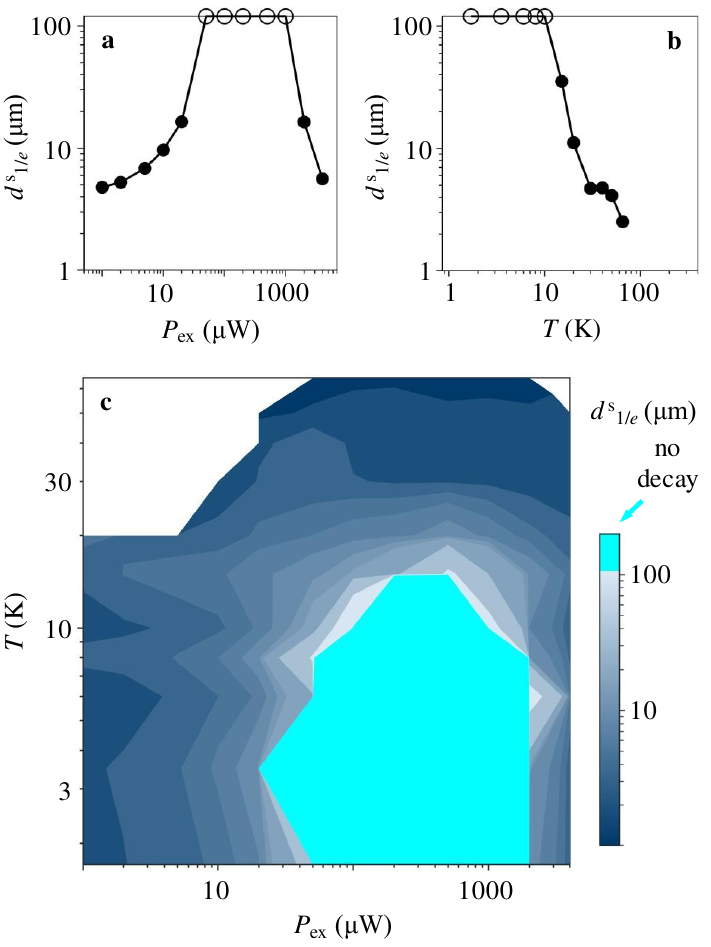}
\caption{Excitation power and temperature dependence of spin transport in LE-IXs.  
(a-c) The $1/e$ decay distance $d^{\rm s}_{1/e}$ of spin density transport $I_{\rm spin} = P n = I_{\sigma^+} - I_{\sigma^-}$ in LE-IXs vs. $P_{\rm ex}$ (a), vs. temperature (b), and vs. $P_{\rm ex}$ and temperature (c). 
$d^{\rm s}_{1/e}$ are obtained from least-squares fitting the LE-IX spin density transport profiles $I_{\rm spin}(x)$ (Fig.~2) to exponential decays in the region $x = 0 - 9$~$\mu$m. The data with the fit indicating diverging $d^{\rm s}_{1/e}$ are presented by circles on the edge (a,b) or by cyan color (c).
The LE-IX spectra are separated from the HE-IX spectra by the gaussian fits. 
The HE-IXs appear in the spectra at high $P_{\rm ex} \gtrsim 0.2$~mW (Fig.~4). 
$T = 3.5$~K (a), $P_{\rm ex} = 0.2$~mW (b).
}
\end{center}
\label{fig:spectra}
\end{figure}

\begin{figure}
\begin{center}
\includegraphics[width=8.5cm]{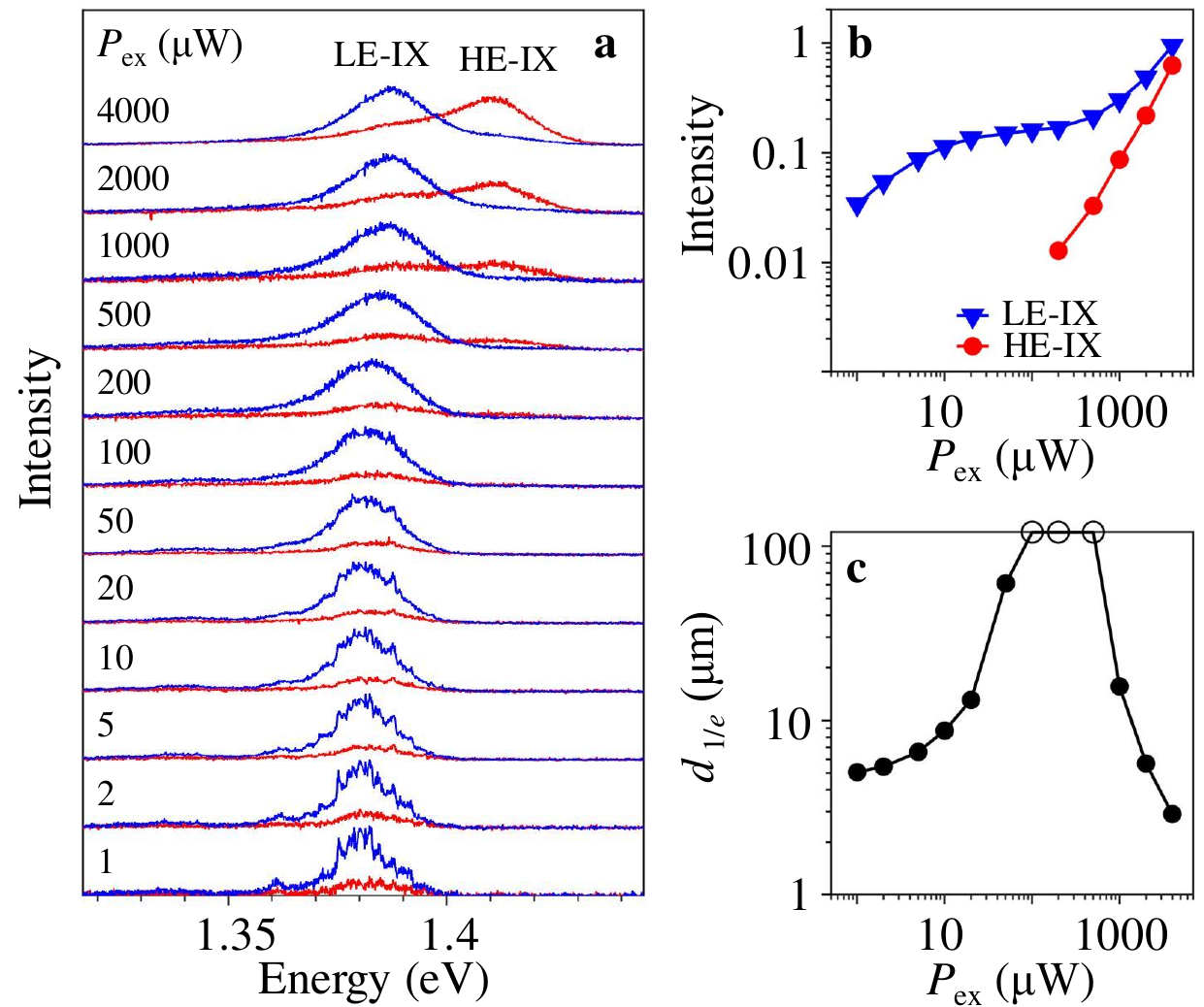}
\caption{Density dependence of IX PL spectra.
(a) The excitation power dependence of co-polarized (blue) and cross-polarized (red) IX spectra. The LE-IX PL is co-polarized. The HE-IX PL is cross-polarized. The HE-IXs appear in the spectra at high $P_{\rm ex} \gtrsim 0.2$~mW. 
The spectral profile separation of LE-IXs and HE-IXs is presented in Fig.~S3 in SI.
(b) The intensity of LE-IX PL (blue triangles) and HE-IX PL (red points) vs. $P_{\rm ex}$. 
(c) The $1/e$ LE-IX transport decay distance $d_{1/e}$ vs. $P_{\rm ex}$. $d_{1/e}$ are obtained from least-squares fitting the spectrally integrated LE-IX PL intensity including both polarizations $I_{\sigma^+} + I_{\sigma^-}$ to exponential decays in the region $x = 0 - 9$~$\mu$m. 
$T = 3.5$~K.
}
\end{center}
\label{fig:spectra}
\end{figure}

The data are discussed below. The spin transport (Fig.~3) is carried by LE-IXs and can be compared with the LE-IX transport (Fig.~4c and Fig.~S7 in SI). Due to the separation $d_z$ between the electron and hole layers, IXs have electric dipoles $ed_z$ and the interaction between IXs is repulsive~\cite{Yoshioka1990}. IXs in moir{\'e} superlattices form a system of repulsively interacting bosons in periodic potentials. The enhancement followed by the suppression of the LE-IX transport with density (Fig.~4c) is in qualitative agreement with the Bose-Hubbard theory of bosons in periodic potentials predicting superfluid at $N \sim 1/2$ and insulating at $N \sim 0$ and $N \sim 1$ phases~\cite{Fisher1989}. For the maximum LE-IX transport distances observed at $P_{\rm ex} \sim 0.2$~mW (Fig.~4c), the LE-IX density $n$ estimated from the energy shift $\delta E$ as $n = \delta E \varepsilon / 4\pi e^2 d_z$~\cite{Yoshioka1990} is $n \sim 2 \times 10^{11}$~cm$^{-2}$ ($d_z \sim 0.6$~nm, the dielectric constant $\varepsilon \sim 7.4$~\cite{Laturia2018}). This density is well below the Mott transition density $n_{\rm Mott} > 10^{12}$~cm$^{-2}$~\cite{Fogler2014, Wang2019}. $N \sim 1/2$ at $n \sim 2 \times 10^{11}$~cm$^{-2}$ for the moir{\'e} superlattice period $b = 17$~nm. This period $b \sim a / \delta \theta$ corresponds to the twist angle $\delta \theta = 1.1^\circ$, which agrees with the angle between MoSe$_2$ and WSe$_2$ edges in the HS (Fig.~S1 in SI). This rough estimate indicates that the observation of LE-IX localization, then long-range transport, and then localization with increasing density (Fig.~4c) is in agreement with the Bose-Hubbard theory~\cite{Fisher1989}. The density dependence of spin transport (Fig.~3a) is qualitatively similar to the density dependence of LE-IX transport (Fig.~4c).

Furthermore, the temperature dependence of spin transport is qualitatively similar to the temperature dependence of LE-IX transport: Both the long-distance spin transport (Fig.~3) and the long-distance LE-IX transport (Fig.~S7 in SI) vanish at $T \sim 10$~K. 

The similarity of the parameter dependence of the spin transport and the IX transport indicates that the dominant mechanism of the decay of spin density is the decay of IX density. The suppression of IX scattering, which leads to the long-distance IX transport (Fig.~S7 in SI), also leads to the long-distance spin transport with $d^{\rm s}_{1/e}$ exceeding 100~$\mu$m and diverging so spin currents show no decay in the HS (Fig.~3). This is consistent with the theory prediction that the suppression of scattering can cause the suppression of spin relaxation~\cite{Dyakonov2008}.

This indicates that the long-distance spin transport with suppressed losses can be achieved at higher temperatures in IX systems with higher superfluidity temperatures $T_{\rm c}$. The theory predicts that the superfluidity temperature for bosons in periodic potentials $T_{\rm c} \sim 4\pi N J$ and higher $T_{\rm c}$ can be achieved in lattices with higher inter-site hopping $J$~\cite{Capogrosso-Sansone2008}. Higher $J$ can be achieved in moir{\'e} superlattices with smaller periods in HS with larger twist angles $\delta \theta$, or in moir{\'e} superlattices with smaller amplitudes that can be realized in HS with the same-TMD electron and hole layers~\cite{Wang2018, Calman2018}, or by lowering the moir{\'e} superlattice amplitude by voltage~\cite{Yu2017, Fowler-Gerace2021}, or by adding a spacer (hBN) layer between the electron and hole layers~\cite{Unuchek2019}. For TMD HS with suppressed moir{\'e} potentials, the theory predicts high-$T_{\rm c}$ superfluidity~\cite{Fogler2014, Berman2017}. This, in turn, can enable the realization of high-temperature long-distance spin transport with suppressed losses.

The above data outline the long-distance spin transport carried by LE-IXs. Figure 4 shows that HE-IX PL appears in the spectrum at high $P_{\rm ex}$. In contrast to the LE-IX PL, which is co-polarized, the HE-IX PL is cross-polarized with the circularly polarized laser excitation. A similar higher-energy IX PL was observed in earlier studies. Various interpretations for multiple IX PL lines were considered, including the excitonic states split due to the conduction band K-valley spin splitting~\cite{Rivera2015}, trions~\cite{Calman2020}, excitonic states indirect in momentum space and split due to the valley energy difference~\cite{Miller2017, Okada2018} or spin-orbit coupling~\cite{Hanbicki2018}, and excitonic states in moir{\'e} superlattice~\cite{Zhang2018, Ciarrocchi2019, Seyler2019, Tran2019, Jin2019, Alexeev2019, Jin2019a}. The data in Fig.~4 agree with the latter and indicate that the HE-IX PL corresponds to two IXs per moir{\'e} cell. The appearance of HE-IX PL in the spectrum (Fig.~4a,b) correlates with the onset of IX transport suppression (Fig.~4c). For the IX transport suppression caused by the high occupation of moir{\'e} cells, outlined above, this correlation indicates that the onset of HE-IX PL corresponds to the appearance of moir{\'e} cells with two IXs per cell. For moir{\'e} cells with two IXs, the IXs can
emit in different polarizations~\cite{Yu2017}, and the intra-cell IX repulsion enhances the IX energy~\cite{HE, lines}. 

In summary, we observed in IXs in a MoSe$_2$/WSe$_2$ HS the long-distance spin transport with $1/e$ decay distances exceeding 100~$\mu$m and diverging so spin currents show no decay in the HS. With increasing IX density, we observed spin localization, then long-distance spin transport, and then reentrant spin localization, in agreement with the Bose-Hubbard theory prediction for superfluid  and insulating phases in periodic potentials due to moir{\'e} superlattices. The suppression of scattering in exciton superfluid suppresses the spin relaxation and enables the long-distance spin transport.

We thank M.M. Fogler and J.R. Leonard for discussions. We especially thank A.H. MacDonald for discussions of IXs in moir{\'e} potentials and A.K. Geim for teaching us manufacturing TMD HS. The studies were supported by DOE Office of Basic Energy Sciences under Award DE-FG02-07ER46449. The HS manufacturing was supported by NSF Grant 1905478.


\begin{references}
\bibitem{HE}
Spin transport in HE-IXs is shorter-range than in LE-IXs and is not considered in this work.
\bibitem{lines}
Narrow PL lines at low densities (Fig.~4a) can be related to localized states and are not considered in this work.
\end{references}
\end{document}


\title{Supporting Information for

Long-distance decay-less spin transport in indirect excitons in a van der Waals heterostructure}

\author{Zhiwen~Zhou}
\affiliation{Department of Physics, University of California at San Diego, La Jolla, CA 92093, USA}
\author{E.~A.~Szwed}
\affiliation{Department of Physics, University of California at San Diego, La Jolla, CA 92093, USA}
\author{D.~J. Choksy}
\affiliation{Department of Physics, University of California at San Diego, La Jolla, CA 92093, USA}
\author{L.~H.~Fowler-Gerace}
\affiliation{Department of Physics, University of California at San Diego, La Jolla, CA 92093, USA}
\author{L.~V.~Butov} 
\affiliation{Department of Physics, University of California at San Diego, La Jolla, CA 92093, USA}

\begin{abstract}
\noindent
\end{abstract}

\maketitle

\renewcommand*{\thefigure}{S\arabic{figure}}

\subsection{The heterostructure details}

The van der Waals MoSe$_2$/WSe$_2$ heterostructure (HS) was assembled using the dry-transfer peel-and-lift technique~\cite{Withers2015}. 
The same HS was used for the studies of IX transport in Ref.~\cite{Fowler-Gerace2023} and the HS manufacturing details are described in Ref.~\cite{Fowler-Gerace2023}. 
The thickness of bottom and top hBN layers is about 40 and 30~nm, respectively. The MoSe$_2$ monolayer is on top of the WSe$_2$ monolayer. The long WSe$_2$ and MoSe$_2$ edges reach $\sim 30$ and $\sim 20$~$\mu$m, respectively, which enables a rotational alignment between the WSe$_2$ and MoSe$_2$ monolayers. The twist angle $\delta \theta = 1.1^\circ$ corresponding to the moir{\'e} superlattice period $b = 17$~nm, which gives $N \sim 1/2$ at the estimated $n \sim 2 \times 10^{11}$~cm$^{-2}$ for the long-distance IX transport as outlined in the main text, agrees with the angle between MoSe$_2$ and WSe$_2$ edges in the HS (Fig.~S1b). 

The accuracies of estimating $\delta \theta$ using the long WSe$_2$ and MoSe$_2$ edges and using SHG are comparable. We do not use SHG for additional estimates of $\delta \theta$ since the intense optical excitation pulses in SHG measurements may cause a deterioration of the HS and may suppress both the long-distance IX transport and the long-distance spin transport. As outlined in the main text, the moir{\'e} potentials can be affected by atomic reconstruction and by disorder and may vary over the HS area.

Figure~S1b presents a microscope image showing the layer pattern of the HS. The layer boundaries are indicated. The hBN layers cover the entire areas of MoSe$_2$ and WSe$_2$ layers. There was a narrow multilayer graphene electrode on the top of the HS around $x = 2$~$\mu$m for $y=0$, Fig.~S1b. This electrode was detached. The IX luminescence reduction around $x =2$~$\mu$m can be related to residual graphene layers on the HS.

\begin{figure}
\begin{center}
\includegraphics[width=10cm]{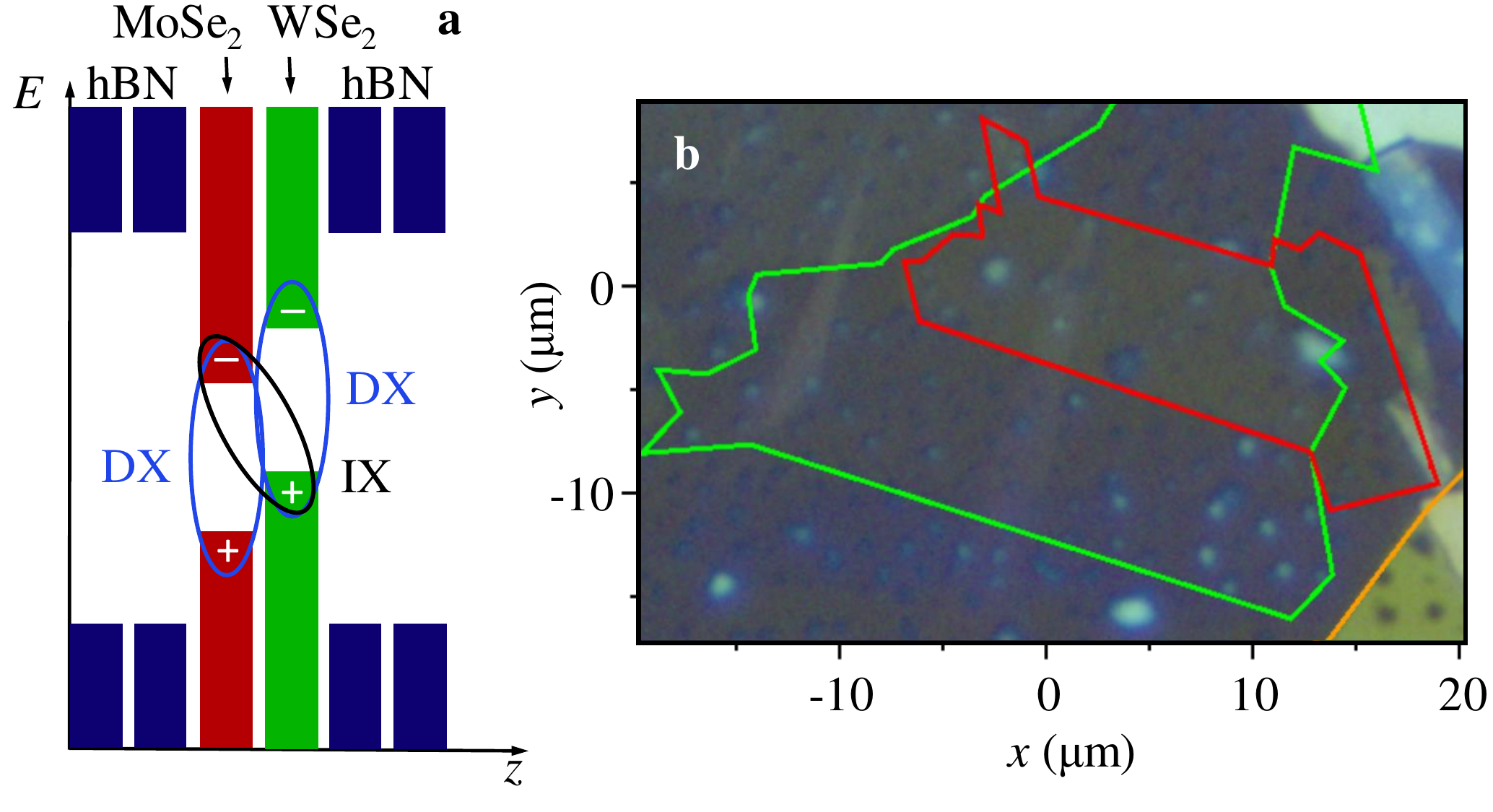}
\caption{(a) Schematic energy-band diagram for the MoSe$_2$/WSe$_2$ HS. The ovals indicate a direct exciton (DX) and an indirect exciton (IX) composed of an electron (–) and a hole (+). (b) A microscope image showing the layer pattern of the HS. The green and red lines indicate the boundaries of WSe$_2$ and MoSe$_2$ monolayers, respectively. The bottom and top hBN layers entirely cover the WSe$_2$ and MoSe$_2$ monolayers, the boundaries of bottom hBN layer are beyond the figure and a part of the boundaries of top hBN layer is shown by the orange line. 
}
\end{center}
\label{fig:spectra}
\end{figure}

We did not verify if stacking in the sample is R or H (AA or AB). This is the subject for future works. We note however, that regardless the stacking type, the data in the paper demonstrate the proof-of-principle for the existence of long-distance spin transport in TMD HS. We note also that the characteristic energies in the IX system in the regime of the long-distance spin transport (Fig.~3), including the IX interaction energy $\delta E \sim 3$~meV and thermal energy $k_{\rm B}T \lesssim 1$~meV ($k_{\rm B}$ is the Boltzmann constant), are considerably smaller than the estimated amplitudes of the moir{\'e} potential for both R and H stacking~\cite{Wu2018, Yu2018, Wu2017, Yu2017}. 
Besides the demonstration of the existence of long-distance spin transport in TMD HS, the studied MoSe$_2$/WSe$_2$ HS allows measuring the density-temperature phase diagram for this phenomenon and comparing it with the theory. However, it is essential to study this phenomenon in other samples and, in particular, to study the dependence on the twist angle.
Both the long-distance IX transport and the long-distance spin transport require the suppression of IX scattering as outlined in the main text and, in turn, require that the disorder in HS is small. It is challenging to manufacture samples with different twist angles, all with sufficiently small disorder, in order to study the dependence of the spin transport on the twist angle and, in turn, on the moir{\'e} superlattice period. Consequently, this remains the subject for future works.

\subsection{Optical measurements}

\begin{figure}
\begin{center}
\includegraphics[width=10.5cm]{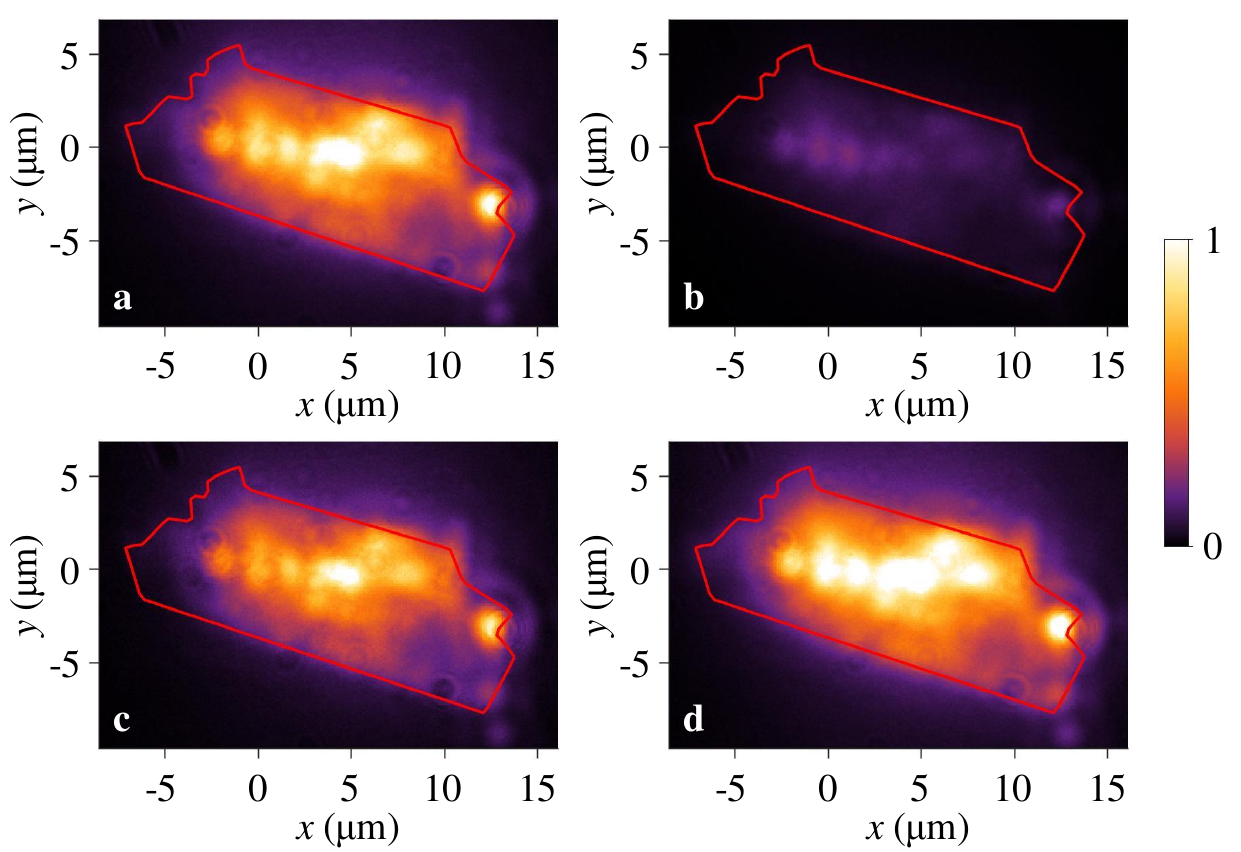}
\caption{(a,b) IX PL images co-polarized $I_{\sigma^+}(x,y)$ (a) and cross-polarized $I_{\sigma^-}(x,y)$ (b) with the circularly polarized laser excitation. The IX PL is selected by a filter $E \lesssim 1.4$~eV.
(c) The spin density image $I_{\rm spin} = P n = I_{\sigma^+} - I_{\sigma^-}$.
(d) The density image $n = I_{\sigma^+} + I_{\sigma^-}$.
The $\sim 2$~$\mu$m laser excitation spot is centered at $(0, 0)$, 
$P_{\rm ex} = 0.2$~mW, $T = 1.7$~K. 
}
\end{center}
\label{fig:spectra}
\end{figure}

Excitons were generated by a cw Ti:Sapphire laser with the excitation energy $E_{\rm ex} = 1.689$~eV. 
PL spectra were measured using a spectrometer with resolution 0.2~meV and a liquid-nitrogen-cooled CCD. The spatial profiles of polarization-resolved IX PL vs. $x$ were obtained from the polarization-resolved PL images detected using the CCD. The signal was integrated from $y = - 0.5$ to $y = + 0.5$~$\mu$m. Figure~S2 shows representative polarization-resolved IX PL images.  

The IX PL kinetics in the sample was measured in earlier studies~\cite{Fowler-Gerace2023} using a pulsed semiconductor laser and a liquid-nitrogen-cooled CCD coupled to a PicoStar HR TauTec time-gated intensifier. These measurements showed that the IX lifetimes in the sample are in the range of $\sim 10 - 30$~ns for the densities and temperatures studied in this work and $\sim 20$~ns for the densities and temperatures corresponding to both the long-distance IX transport and the long-distance spin transport carried by IXs. These measurements also showed that for the densities and temperatures corresponding to both the long-distance IX transport and the long-distance spin transport carried by IXs, the IX propagation is characterized by the average velocity of the IX cloud expansion $v \sim \Delta R/ \Delta t \sim 5 \times 10^4$~cm/s. Since the spin transport is carried by IXs, the IX transport kinetics can give a rough estimate for the spin transport kinetics, however, spatially- and polarization-resolved imaging experiments are needed to measure the spin transport kinetics in IXs and this is the subject for future works. 

The experiments were performed in a variable-temperature 4He cryostat. The sample was mounted on an Attocube xyz piezo translation stage allowing adjusting the sample position relative to a focusing lens inside the cryostat. 
All physical phenomena presented in this work are reproducible after ca.~100 cooling down to 2~K and warming up to room temperature.

\subsection{LE-IX and HE-IX spectral profile separation}

We used two methods to separate LE-IX spectra from HE-IXs spectra: (i) by the spectral integration in the range $E < 1.4$~eV where LE-IXs dominate the spectra (Fig.~1a and 4a) and (ii) by the gaussian fits (Fig.~S3b). These two methods give similar results: compare Fig.~1b,c with Fig.~S4a,b, Fig.~2a-c with Fig.~S5a-c, and Fig.~3a-c with Fig.~S6a-c where the LE-IX spectra were separated from HE-IX spectra using these two different methods in the two figures in each pair of the figures. 

\begin{figure}
\begin{center}
\includegraphics[width=15cm]{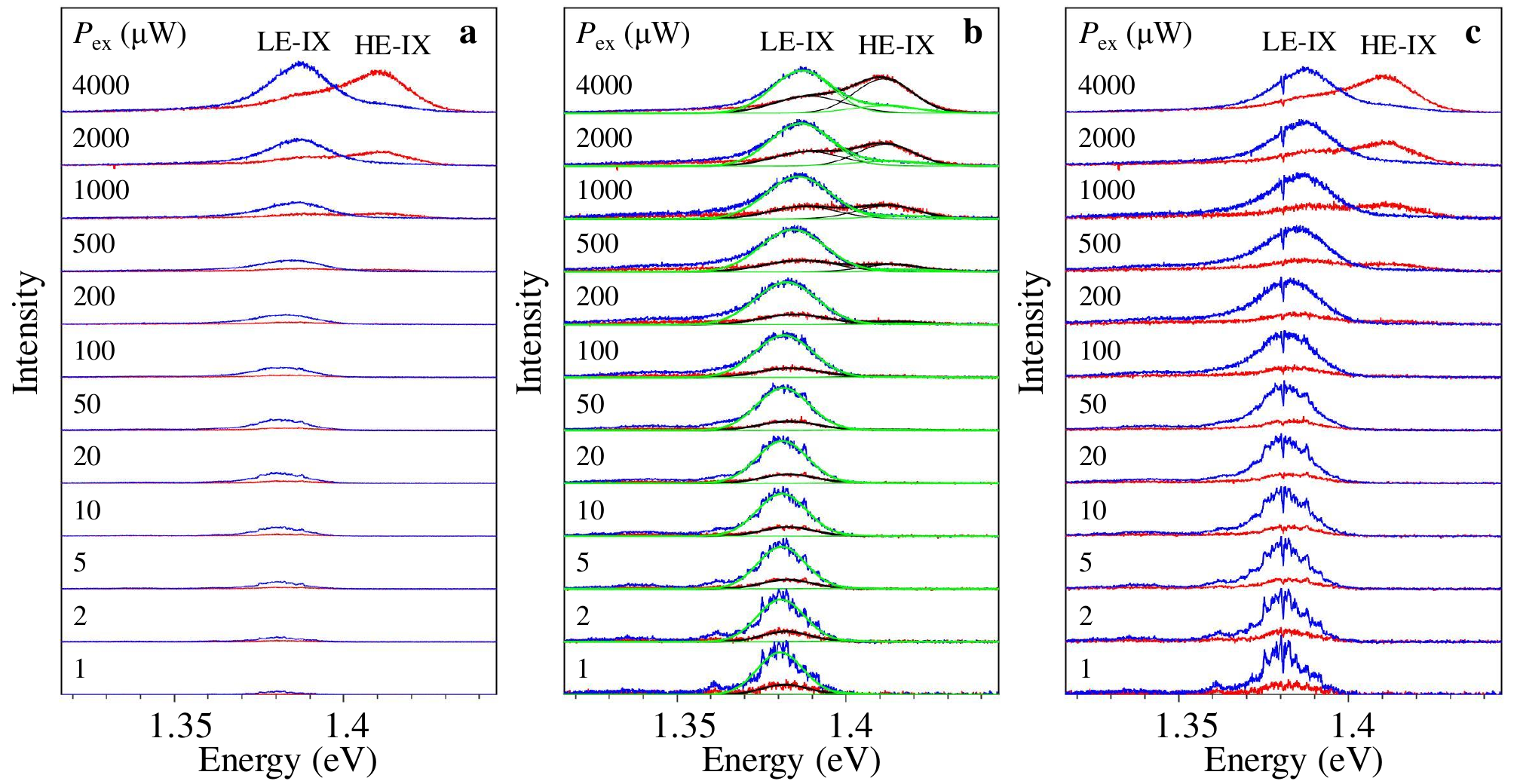}
\caption{(a) The excitation power dependence of co-polarized (blue) and cross-polarized (red) IX spectra. The LE-IX PL is co-polarized. The HE-IX PL is cross-polarized. The HE-IXs appear in the spectra at high $P_{\rm ex} \gtrsim 0.2$~mW. The same spectra with normalized intensities are shown in Fig.~4a. 
(b) The same spectra as in Fig.~4a with the spectral profile separation of the co-polarized and cross-polarized PL of LE-IXs and HE-IXs. The gaussian fits to the co-polarized (cross-polarized) LE-IX spectra and HE-IX spectra are shown by the thin green (black) lines. The sum of the gaussians shown by the thin green (black) lines is shown by the thick green (black) line. 
(c) The same spectra as in (b) without excluding the part of the spectra affected by the CCD defect at $\sim 1.38$~eV, which causes an intensity reduction. The parts of the spectra affected by this defect are excluded from the spectra in (b) and other spectra in the paper. 
$T = 3.5$ K.
}
\end{center}
\label{fig:spectra}
\end{figure}

\begin{figure}
\begin{center}
\includegraphics[width=10cm]{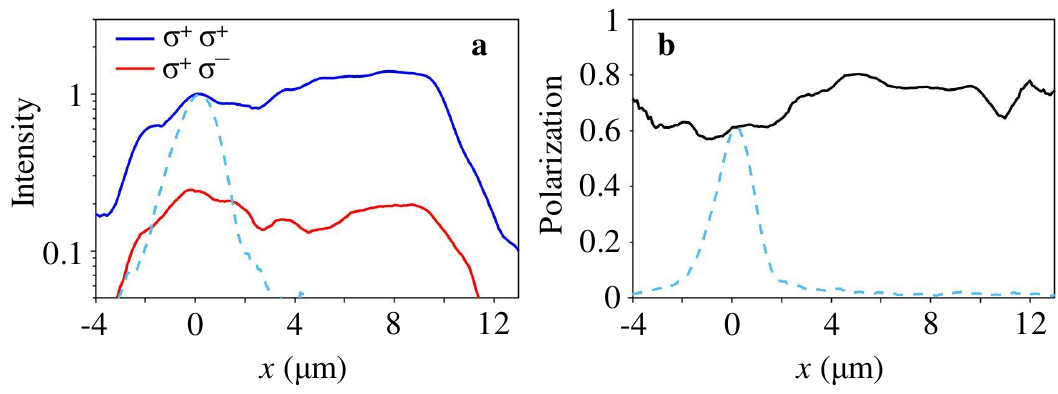}
\caption{Same as Fig.~1b,c for LE-IXs with the LE-IX spectra separated from the HE-IX spectra by the gaussian fits. Selecting LE-IXs by the spectral integration in the range $E < 1.4$~eV (as in Fig.~1b,c) or by the spectral integration of the gaussian fits (as in Fig.~S4a,b) give similar results.
}
\end{center}
\label{fig:spectra}
\end{figure}

\begin{figure}
\begin{center}
\includegraphics[width=15cm]{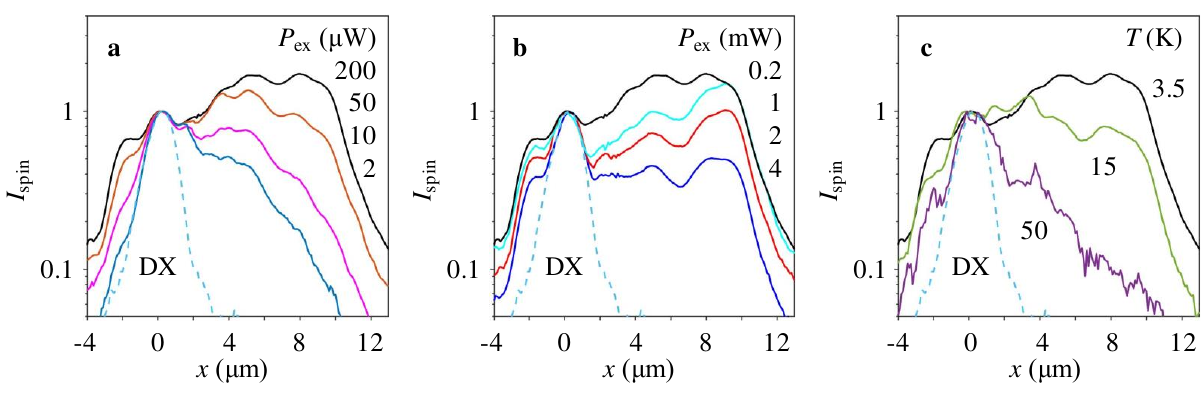}
\caption{Same as Fig.~2 for the LE-IXs with the LE-IX spectra separated from the HE-IX spectra by the gaussian fits. Selecting LE-IXs by the spectral integration in the range $E < 1.4$~eV (as in Fig.~2) or by the spectral integration of the gaussian fits (as in Fig.~S5) give similar results.
}
\end{center}
\label{fig:spectra}
\end{figure}

\begin{figure}
\begin{center}
\includegraphics[width=8.5cm]{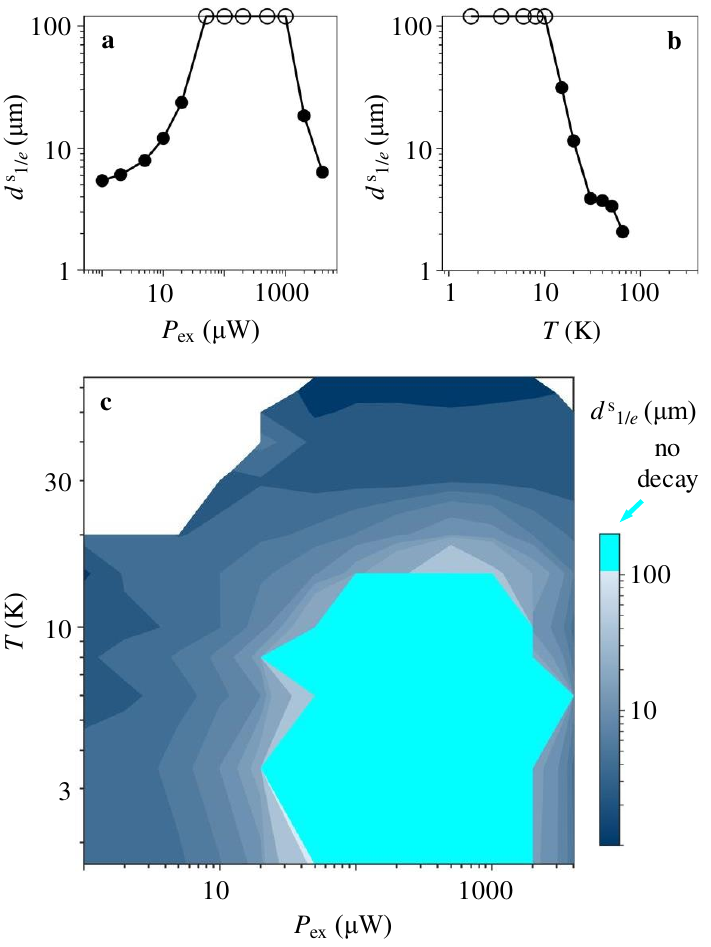}
\caption{Same as Fig.~3 for LE-IXs with the LE-IX spectra separated from the HE-IX spectra by the spectral integration in the range $E < 1.4$~eV. Selecting LE-IXs by the spectral integration in the range $E < 1.4$~eV (as in Fig.~S6) or by the spectral integration of the gaussian fits (as in Fig.~3) give similar results.
}
\end{center}
\label{fig:spectra}
\end{figure}

\subsection{The density and temperature dependence of LE-IX transport}

\begin{figure}
\begin{center}
\includegraphics[width=8.5cm]{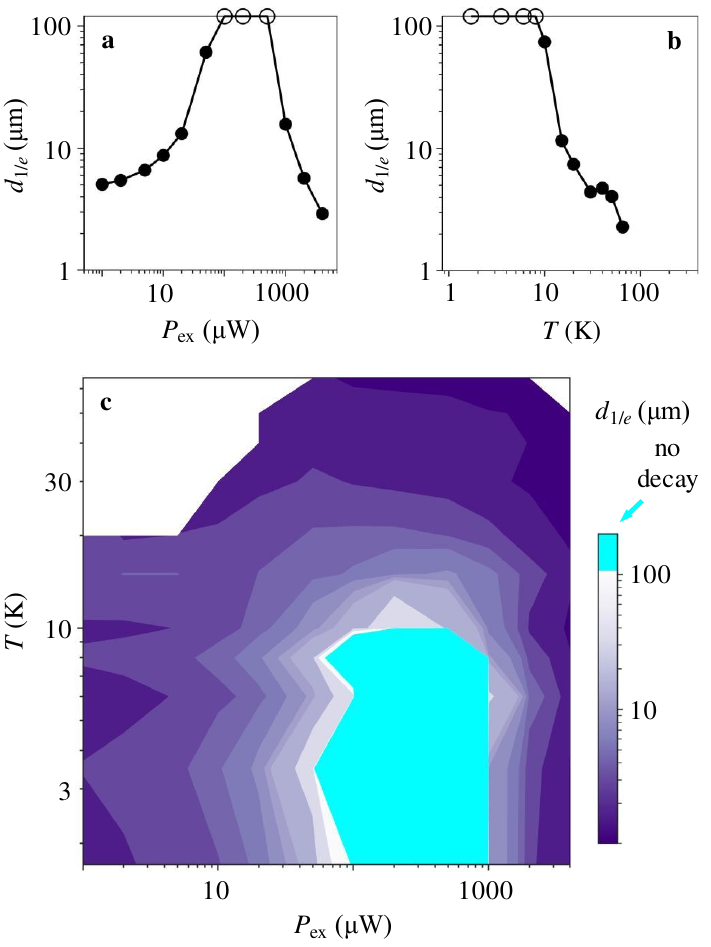}
\caption{Excitation power and temperature dependence of LE-IX transport. 
(a,b) The $1/e$ decay distance $d_{1/e}$ of LE-IX PL intensity $I = I_{\sigma^+} + I_{\sigma^-}$ vs. $P_{\rm ex}$ (a), vs. temperature (b), and vs. $P_{\rm ex}$ and temperature (c). 
$d_{1/e}$ are obtained from least-squares fitting the LE-IX intensity profiles $I(x)$ to exponential decays in the region $x = 0 - 9$~$\mu$m. The data with the fit indicating diverging $d_{1/e}$ are presented by circles on the edge (a,b) or by cyan color (c).
The LE-IX spectra are separated from the HE-IX spectra by the spectral integration of the gaussian fits.
$T = 3.5$~K (a), $P_{\rm ex} = 0.2$~mW (b).
}
\end{center}
\label{fig:spectra}
\end{figure}

\begin{figure}
\begin{center}
\includegraphics[width=8.5cm]{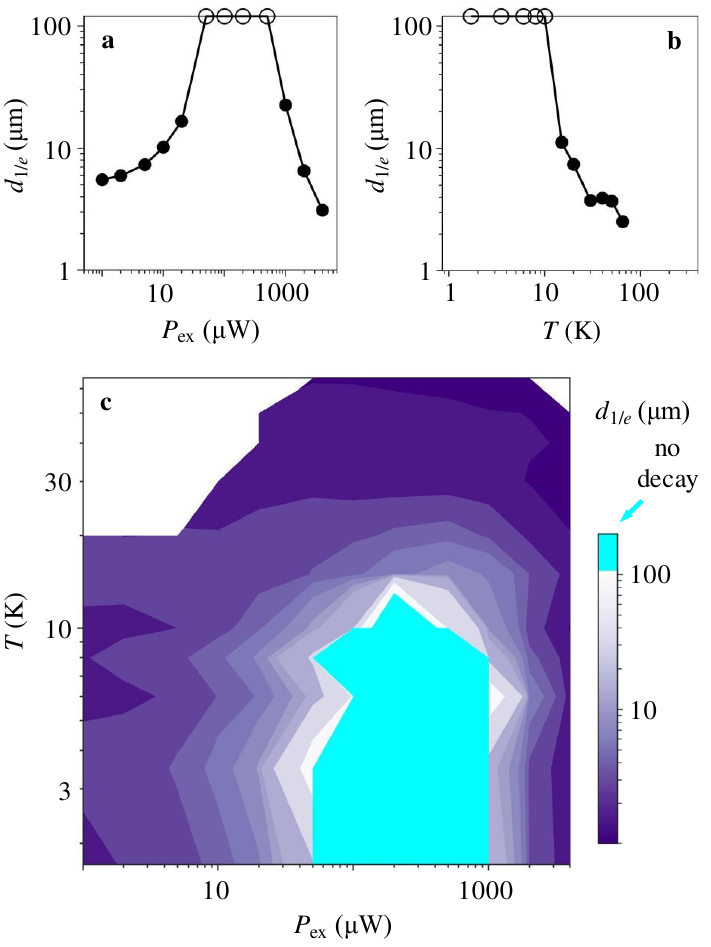}
\caption{Same as Fig.~S7 for LE-IXs with the LE-IX spectra separated from the HE-IX spectra by the spectral integration in the range $E < 1.4$~eV. Selecting LE-IXs by the spectral integration in the range $E < 1.4$~eV (as in Fig.~S8) or by the spectral integration of the gaussian fits (as in Fig.~S7) give similar results.
}
\end{center}
\label{fig:spectra}
\end{figure}

The LE-IX transport is characterized by the $1/e$ decay distance $d_{1/e}$ of the LE-IX PL intensity $I = I_{\sigma^+} + I_{\sigma^-}$ (Fig.~S7 and S8). In these two figures LE-IX spectra are separated from HE-IX spectra using the two methods outlined above. Figures~S7 and S8 show that these two methods give similar results. 

The density and temperature dependence of LE-IX transport (Figs.~S7 and S8) is qualitatively similar to the density and temperature dependence of IX transport~\cite{Fowler-Gerace2023}. The latter refers to transport of all IXs, including LE-IXs and HE-IXs, and has a narrower $d_{1/e}(P_{\rm ex})$ profile, in particular, because HE-IXs have shorter transport distances. The density and temperature dependence of spin transport in LE-IXs (Fig.~3) is qualitatively similar to the density and temperature dependence of LE-IX transport (Figs.~S7 and S8) as outlined in the main text.

The LE-IX intensity enhancement with density slows at intermediate densities (Fig.~4b). This is qualitatively consistent with a more effective LE-IX cloud expansion from the excitation spot due to the longer-distance LE-IX transport at these densities (Figs.~S7 and S8).

\subsection{The temperature dependence of co-polarized and cross-polarized IX PL}

The temperature dependence of co-polarized and cross-polarized IX spectra is presented in Fig.~S9. The temperature increase causes a reduction of the degree of circular polarization of LE-IX PL $P = (I_{\sigma^+} - I_{\sigma^-})/(I_{\sigma^+} + I_{\sigma^-})$ (Fig.~S9c). However, this reduction is rather small, with no sharp changes at $T \sim 10$~K where the transport of spin polarization density $I_{\rm spin}$ carried by LE-IX sharply drops (Fig.~3). This confirms that the dominant mechanism of the decay of spin density $I_{\rm spin} = Pn$ is the decay of IX density as outlined in the main text.

\begin{figure}
\begin{center}
\includegraphics[width=15cm]{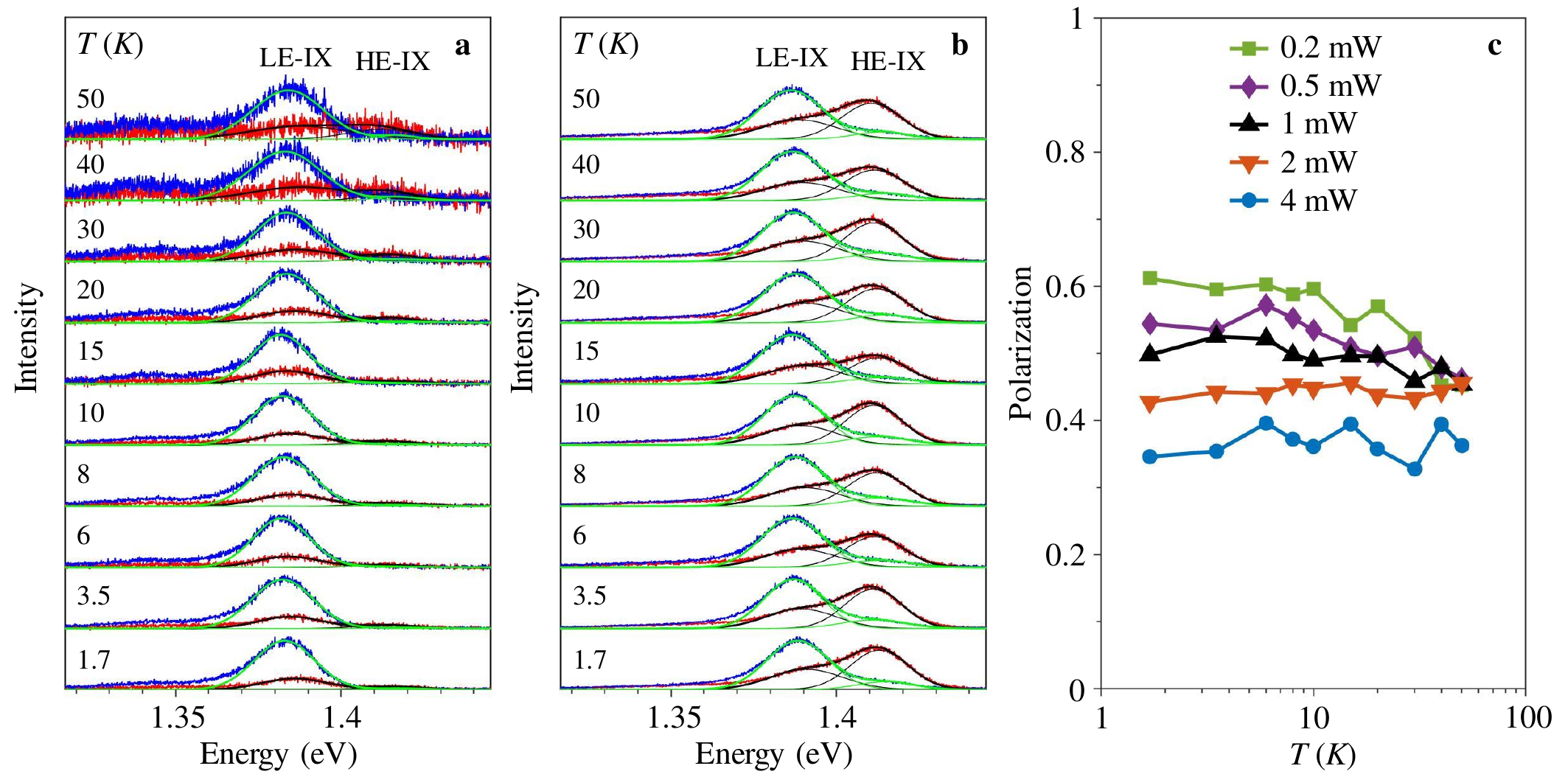}
\caption{(a,b) The temperature dependence of co-polarized (blue) and cross-polarized (red) IX spectra at $P_{\rm ex} = 0.2$~mW (a) and 4~mW (b). The LE-IX PL is co-polarized. The HE-IX PL is cross-polarized. The HE-IXs appear in the spectra at high $P_{\rm ex} \gtrsim 0.2$~mW. The intensities are normalized. (c) The degree of circular polarization of LE-IX PL $P = (I_{\sigma^+} - I_{\sigma^-})/(I_{\sigma^+} + I_{\sigma^-})$ vs. temperature at different $P_{\rm ex}$. The LE-IX spectra are separated from the HE-IX spectra by the gaussian fits.
}
\end{center}
\label{fig:spectra}
\end{figure}